\documentclass[letterpaper,amsmath,amssymb,showpacs,prl]{revtex4}
\pdfoutput=1
\usepackage{graphicx}% Include figure files
\usepackage{dcolumn}% Align table columns on decimal point
\usepackage{bm}% bold math
\usepackage[usenames,dvipsnames]{color}
\usepackage{textcomp}
\usepackage{array}% http://ctan.org/pkg/array
\newcolumntype{M}{>{\centering\arraybackslash}m{\dimexpr.25\linewidth-2\tabcolsep}}

\newcommand{\betsgacl}{$\lambda$-(BETS)$_2$GaCl$_4$}

\newcommand{\etpoly}{$\beta^{\prime\prime}$-(ET)$_2$SF$_5$CH$_2$CF$_2$SO$_3$}

\newcommand{\cuncs}{$\kappa$-(ET)$_2$Cu(NCS)$_2$}
\newcommand{\etnhfour}{$\alpha$-(ET)$_2$NH$_4$(SCN)$_4$}

\newcommand{\drop}[1]{\relax}

\newcommand{\note}[1]{\relax}

\renewcommand{\section}[1]{\relax}
\begin{document}

%\preprint{APS/123-QED}

\title{Radial modulation enhances critical fields in 2d superconductors}% Force line breaks with \\

\author{William A.\ Coniglio}
\affiliation{National High Magnetic Field Laboratory, Tallahassee, Florida 32310}
\author{Charles C.\ Agosta}%
\affiliation{Department of Physics, Clark University, Worcester, Massachusetts 01610}%

\date{\today}% It is always \today, today,

\begin{abstract}
The 1964 reports of Fulde, Ferrell, Larkin, and Ovchinnikov (FFLO or LOFF) on paramagnetic enhancement of superconductivity suggested that superconductivity can persist at applied magnetic fields above both its orbital and paramagnetic limits. By forming spatially alternating superconducting and paramagnetic regions, the increase in local magnetic field in the paramagnetic region allows a reduction in field inside the superconductor. We present an FFLO phase diagram model for layered organic superconductors and confirm it with high magnetic field data from four materials. Our work suggests that paramagnetic and superconducting regions form as radially alternating rings about each vortex rather than plane waves, as FFLO is usually described.\end{abstract}

\maketitle
%{
%\renewcommand{\numberline}[1]{\relax}
%\tableofcontents
%}
%introduction: big bang, pretty picture
%
%[chuck section]
%
Fifty years ago, an ordered coexistence of superconductivity and paramagnetism was predicted independently by theorists Fulde and Ferrell\cite{ff_1964} and Larkin and Ovchinnikov\cite{lo_1965}. As interest in the coexistence of magnetism and superconductivity grows, the impetus to experimentally realize long-range ordered states such as FFLO has increased.\cite{beyer_wosnitza_jltp_2013,aperis_prl_2010,koutroulakis_prl_2010,radovan_fortune03_nature,bianchimovshovich03} In this Letter, we link recent experimental results with theory by proposing and testing a model for FFLO behavior in organic superconductors. The model accurately predicts the long-observed enhanced critical field and intermediate phase line that is observed at low temperature in several materials and yields valuable information about the fundamental limit of superconductivity in high magnetic fields.

%Two challenges lie at the heart of the experimental search for FFLO superconductors: First, it would be efficient optimize our choice of sample based on the expected symmetry and characteristics of the ordering, but there are promising FFLO theories for many cases. Reports suggest the phase exists in both organic and heavy fermion superconductors, but the details for each class likely differ; here, we focus on the organics. Second, the experimental conditions necessary to reach the FFLO phase space in these samples are daunting: temperatures well below one Kelvin, magnetic fields of 10--35 Tesla, and rotation resolution smaller than a degree. Since connections between theory and experiment are not well established, careful investigation requires many hours of cryostat time and thousands of field sweeps.

Organic superconductors serve as chemically tuneable model systems for many superconducting phenomena. They are crystallographically clean, one- or two-dimensional, and highly anisotropic. The two-dimensional ones are grown as alternating anion and organic cation layers. The cation layer conducts hundreds of times better than the anion layer and possesses almost all the superconducting Cooper pairs. When we orient the magnetic field along the layers, the vortices (magnetic field quanta) choose a path through the poorly conducting anion layer, a phenomenon known as the vortex lock-in effect.\cite{lee_chaikin02_prl} Vortex lock-in provides the crucial mechanism to reduce the effect of orbitals that limit the critical fields of most superconductors, making it a desirable property both for FFLO research and developing practical low-dimensional superconductors.

Adding magnetic field ordinarily aligns electron spins, but these electrons are bound as superconducting Cooper pairs, which break down at the Chandrasekhar-Clogston (sometimes known as Pauli) paramagnetic limit and result in the destruction of superconductivity.\cite{clogston62_prl,chandrasekhar_1962} FFLO ordering offers a compromise. If some of the Cooper pairs could be broken and their spins aligned with the field, the reduction in free energy would permit the remaining Cooper pairs to superconduct under yet higher magnetic fields. A major topic in FFLO theory is therefore finding which symmetry provides the lowest free energy for the superconducting and paramagnetic regions to share space in the sample.\cite{denisov_buzdin_shimahara_prb_2009}

\begin{figure}
\begin{center}
\includegraphics[width=0.4\textwidth]{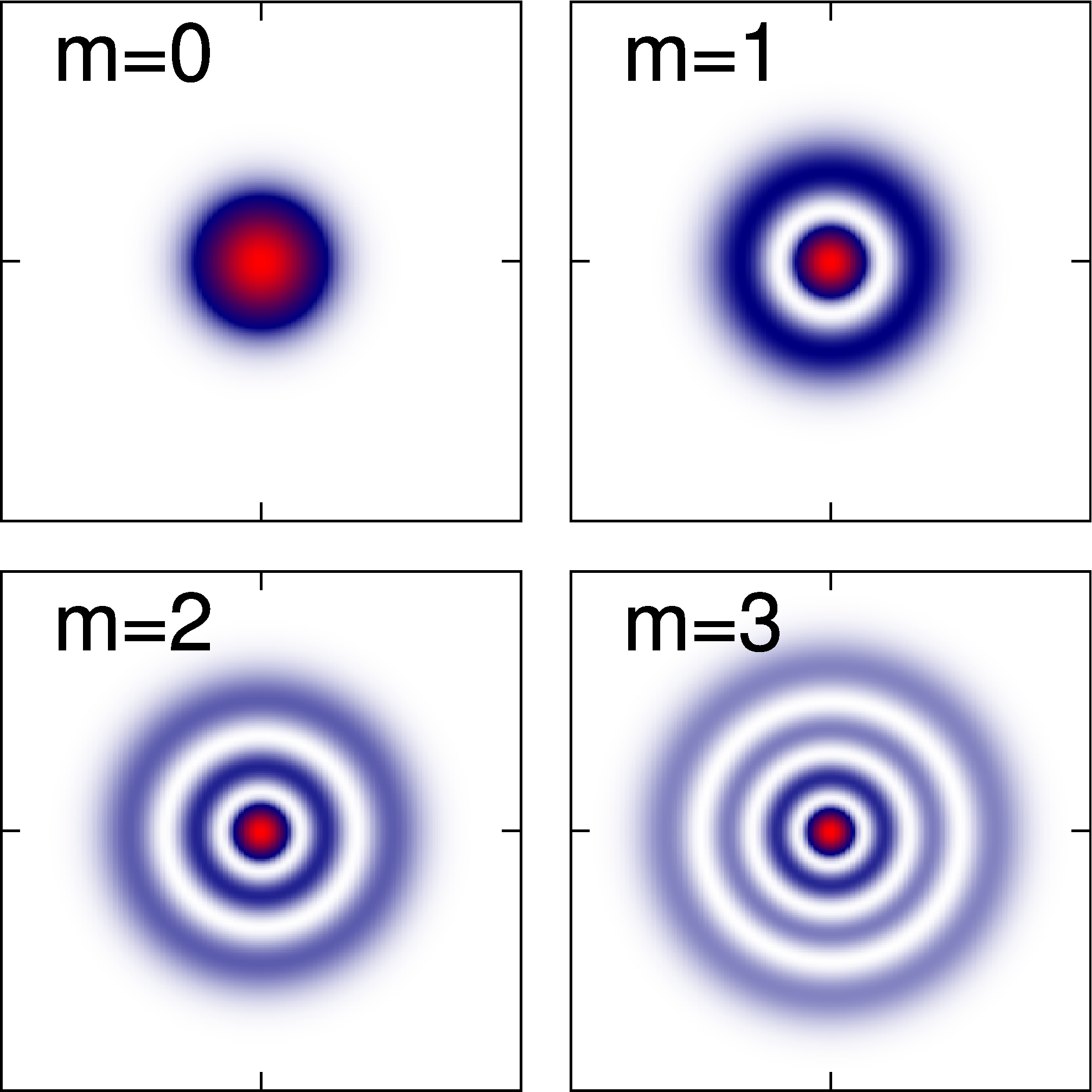}
\end{center}
\caption{\label{fig:vortices}(Color online) Spatial extent of FFLO vortex wave functions proposed by Buzdin and Brison. White is superconducting; red at the center is normal; blue/gray is intermediate. Each plot is $10\xi_\parallel$ across.}
\end{figure}
Theoretical studies of FFLO have proceeded continuously, and there is a large body of excellent work on the topic.\cite{shimahara_prb_2009} From an experimental point of view, the challenge is to first identify samples that exhibit the phenomena, measure them, and match their behavior with appropriate theory that is computationally realizable. We found validity and computational tractability in a model by Buzdin and Brison (BB).\cite{buzdin_brison_1996,houzet_buzdin_bulaevskii_prl_2002} By an appropriate choice of gauge, they consider both paramagnetic and orbital limiting using a basis set that corresponds to vortices containing multiple flux quanta. Each resulting vortex is radially modulated. The wavefunctions resemble ripples around a pebble thrown into a pond, as shown in Fig.~\ref{fig:vortices}.

\section{Formulation of the model}
%Buzdin-Brison vortices provide the correct form of FFLO ordering in this case, but 
Buzdin and Brison follow the work of Werthamer, Helfand, and Hohenberg (WHH). The WHH model appears in Eq.~28 of their classic 1966 paper.\cite{werthamer_helfand_1966} The Green's functions of the two models differ slightly. Buzdin and Brison include Zeeman and multiple quantum orbital terms, while WHH include Zeeman, single quantum orbital, and spin-orbit scattering terms. Our first challenge is to combine them in such a way that the ground state of our new model agrees with the well-established WHH model, which we have shown to apply to these systems.\cite{coniglio_bets_2011} We will also insist on using the same fit parameters as WHH. Two variables scale the temperature and field axes, and two determine the shape of the phase diagram. While the parameter set seems large, each one has physical meaning and is required to realize the range of phase diagrams found in the organic superconductors.

To reconcile the Buzdin-Brison ground state with the WHH model, we find that there must be two sources of orbital limiting when the applied field is oriented parallel to the conducting layers. 1) Due to the vortex lock-in effect,\cite{lee_chaikin02_prl} parallel vortices form between the planes and spill slightly into the highly conducting portion. This form of orbital limiting is treated successfully by a term in the Green's function of the WHH model, which was not included in the original BB formula. We restore it in Eq.~\ref{eq:fflocleanlimit} as $(1-\alpha\lambda_{so})\bar{h}/t$, with the subtraction accounting for the portion of orbital limiting described next. 2) There must be through-plane vortices, which are treated explicitly by the BB model and give rise to FFLO behavior. Given the geometry of the system, we find ourselves curious about the source of these vortices. A clue comes from comparison of the effects of spin-orbit scattering in the WHH model and orbitals in the BB model. BB orbitals in the ground state may be reproduced in the WHH model by introducing an appropriate amount of orbital limiting ($1/\alpha$) and spin-orbit scattering ($\lambda_{so}$). We suggest that spin-orbit scattering is responsible for through-plane vortices. After some algebra and care with units, we recover the final term in the denominator of Eq.~\ref{eq:fflocleanlimit}.

After constructing Eq.~\ref{eq:fflocleanlimit} in the above manner, the ground state in our model reproduces the WHH phase diagram over its entire parameter space. Given the general familiarity with  WHH, we stay with the same notation, normalization, and method of discontinuity removal and leave further refinements for future work. The four parameters to our model are exactly those that appear in WHH, and we stress that the $m=0$ phase here matches the WHH phase diagram exactly. Since two of the parameters are temperature and field, the phase diagram shape is entirely controlled by the two remaining parameters, $\alpha$ and $\lambda_{so}$. All of the examples in Fig.~\ref{fig:fflophases} are produced by combinations of these two parameters.
\begin{widetext}
\begin{equation}
\label{eq:fflocleanlimit}
\ln\frac{1}{t} = \sum_{n=0}^{\infty}
	\frac{2}{\left|2n+1\right|}
	 - \displaystyle\int_0^{\infty}
	\textnormal{Re} \left\{ \frac{(-1)^m L_m(x) \exp(-x/2) \,dx}
		{\sqrt{ \left[ \left|2n+1\right| + \left(1-\alpha\lambda_{so}\right)\bar{h}/t + i\alpha\bar{h}/t\right]^2
		+ x\alpha\lambda_{so}\bar{h}/t^2 }}\right\}
\end{equation}
\end{widetext}

As discussed by Buzdin and Brison, $L_m$ are the Laguerre polynomials $L^k_m$ with $k=0$. The index $m$ indicates the Landau level of the vortex. At any point in $T$-$H$ phase space, the superconductor should be in the lowest indexed phase whose transition line lies above the given point. When it exhausts all superconducting phases at $H_{c2}$, it becomes a normal metal.

We may make one final adjustment to account for normal electron scattering. FFLO superconductivity requires an exceptionally clean material whose mean free path is several times longer than the coherence length. As an approximate treatment for finitely clean samples, we truncate the integration at the normalized mean free path of the material, $r^2=\ell^2/\xi_\parallel^2$, where $\ell$ is the mean free path, and $\xi_\parallel$ is the in-plane superconducting coherence length. We normalize the wavefunction by dividing the integral by $1-\exp(-r^2/2)$. From this approximation, we deduce that the FFLO phase becomes possible when $\ell > 3.5\xi_\parallel$.

The model in Eq.~\ref{eq:fflocleanlimit} may be computed in seconds using a nonlinear root-finding algorithm. It is written in a form that is convenient for computation if care is taken to ensure the infinite sum and integration are carried out over a sufficiently large interval with high precision. We encourage others in the field to implement it themselves.
%When modeling the phase diagram of real materials, the mean free path may be categorized in three ways. 1) For $r>7$, the material is practically in the clean limit. The first five Landau levels are unperturbed by impurities, and the normalizing denominator is nearly unity. 2) For the middle case, $3.5 < r < 7$, the truncated form should be used with the appropriate value of $r$ and normalizing denominator, as the FFLO phase will be somewhat suppressed by the finite mean free path. 3) When $r < 3.5$, the FFLO phase is not possible, and the standard WHH model should be used. There is no discontinuity at the boundaries between model behavior, as this FFLO model builds upon WHH.
\newlength\colwidth
\colwidth=0.75in
\begin{figure}\begin{center}
\begin{tabular*}{4\colwidth}{m{\colwidth}m{\colwidth}m{\colwidth}m{\colwidth}}
\centering$\alpha=4$ & \centering$\alpha=8$ & \centering$\alpha=16$ & ~ \\
\includegraphics[width=\colwidth]{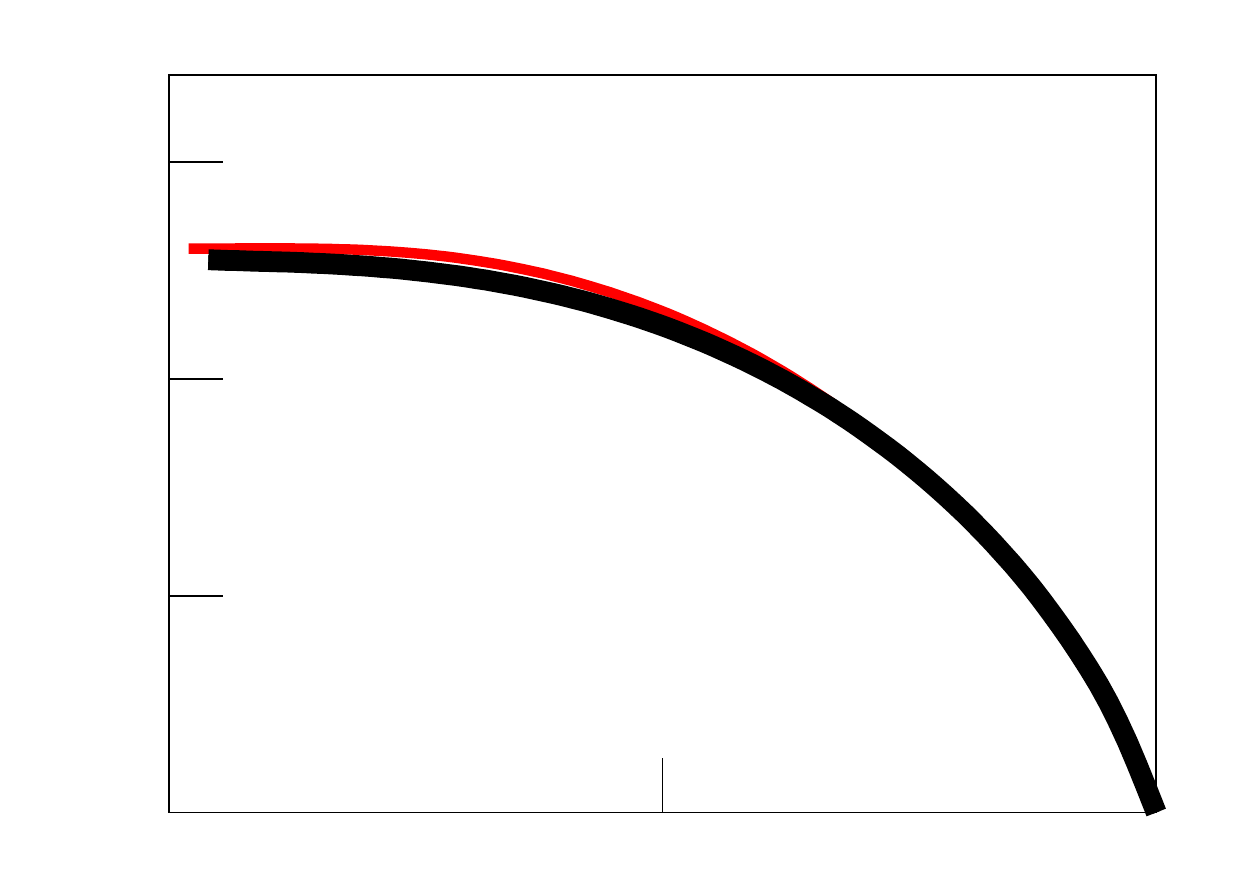} &
\includegraphics[width= \colwidth]{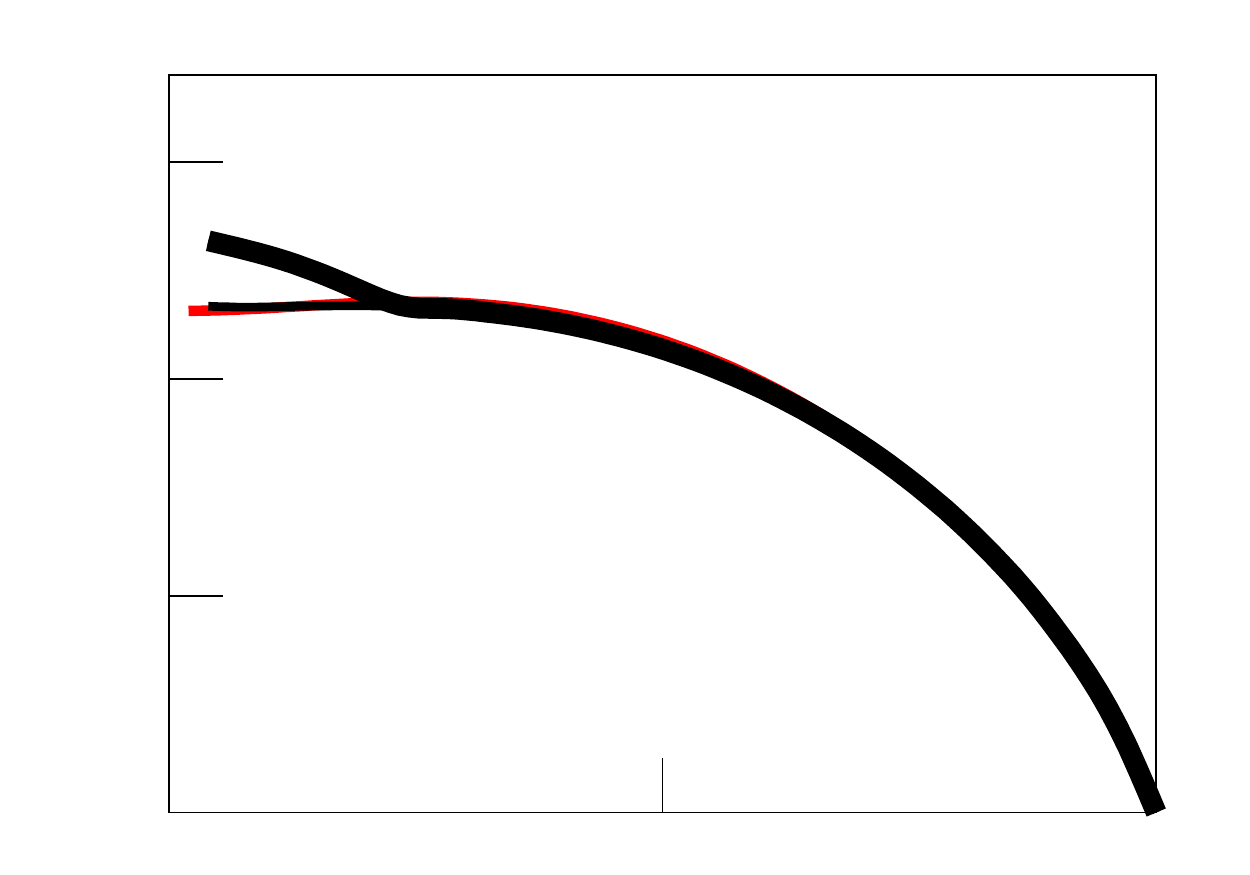} &
\includegraphics[width= \colwidth]{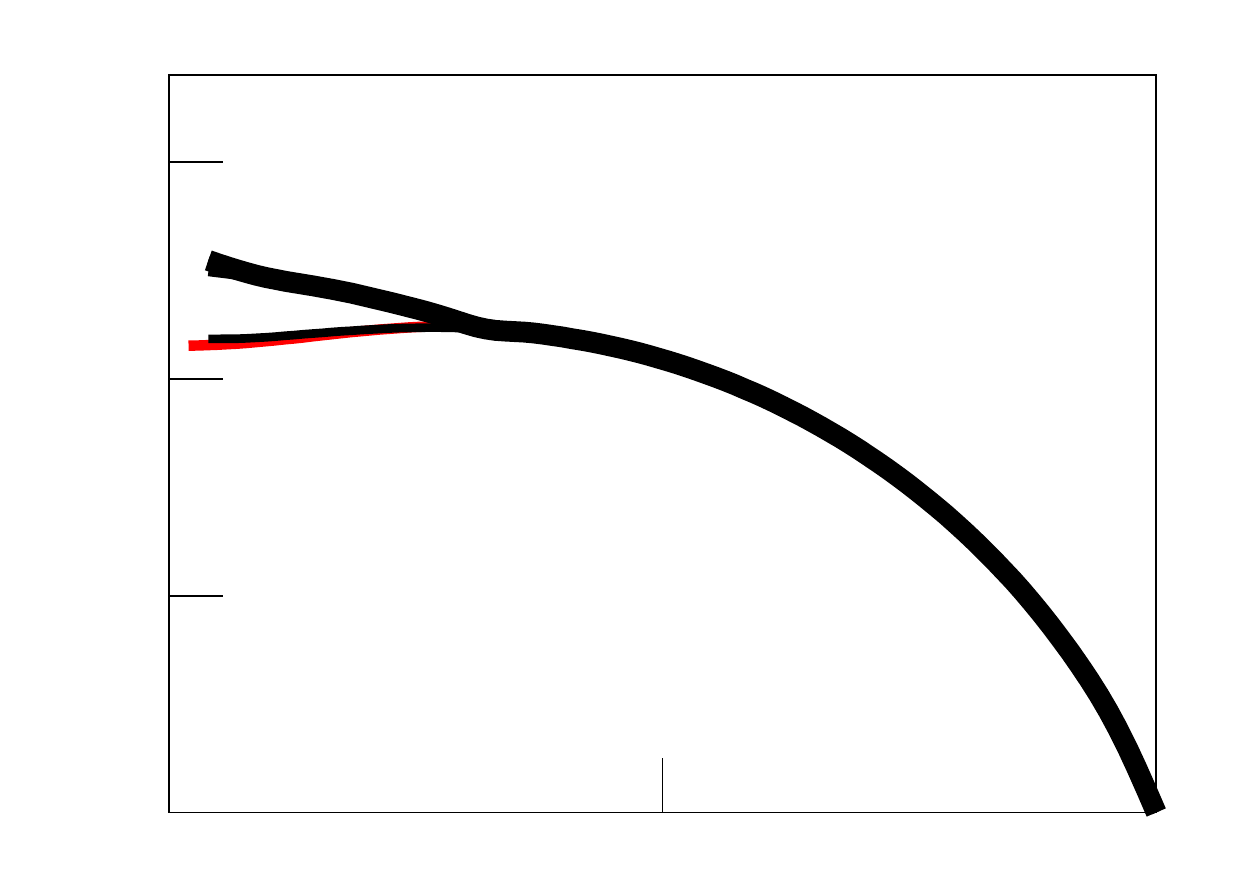} &
$\lambda_{so}=1/\alpha$ \\
\includegraphics[width= \colwidth]{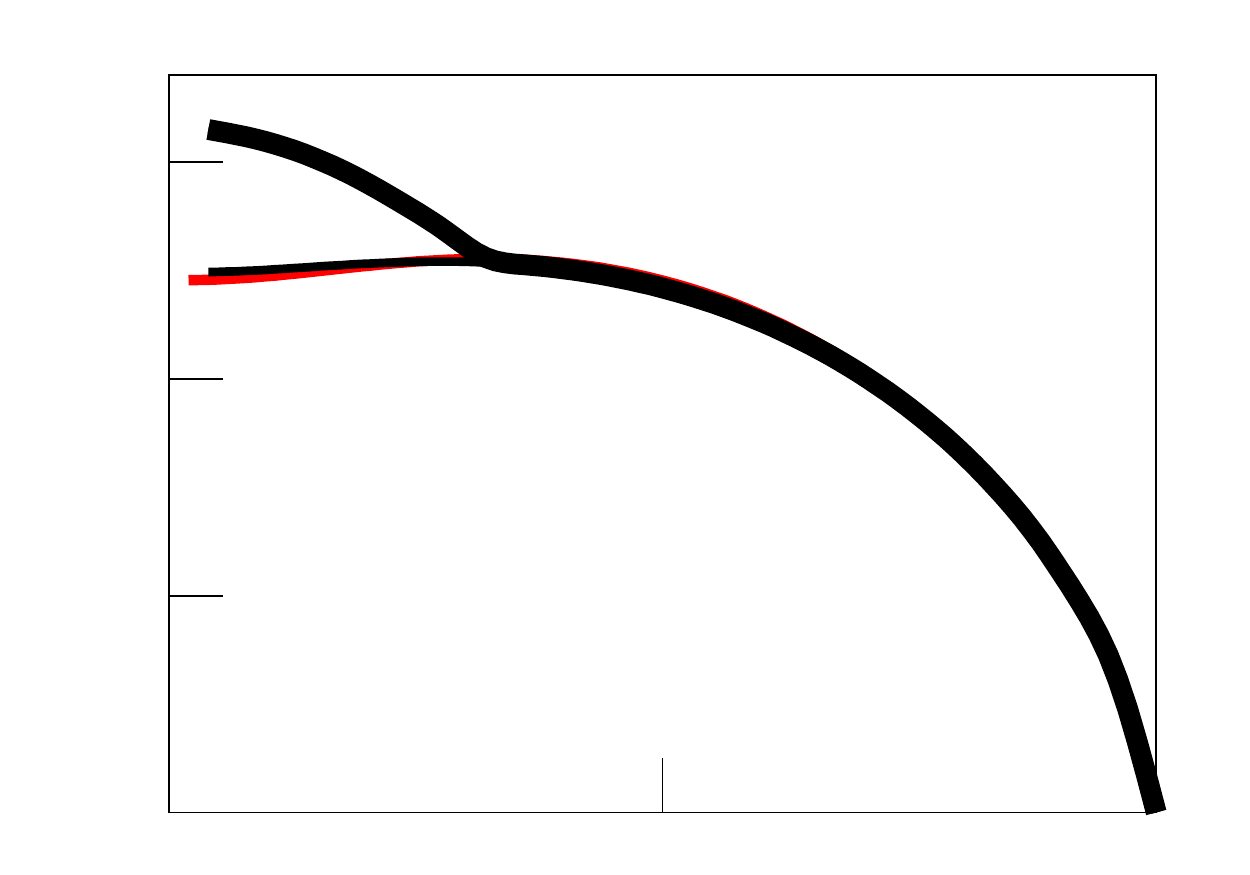} &
\includegraphics[width= \colwidth]{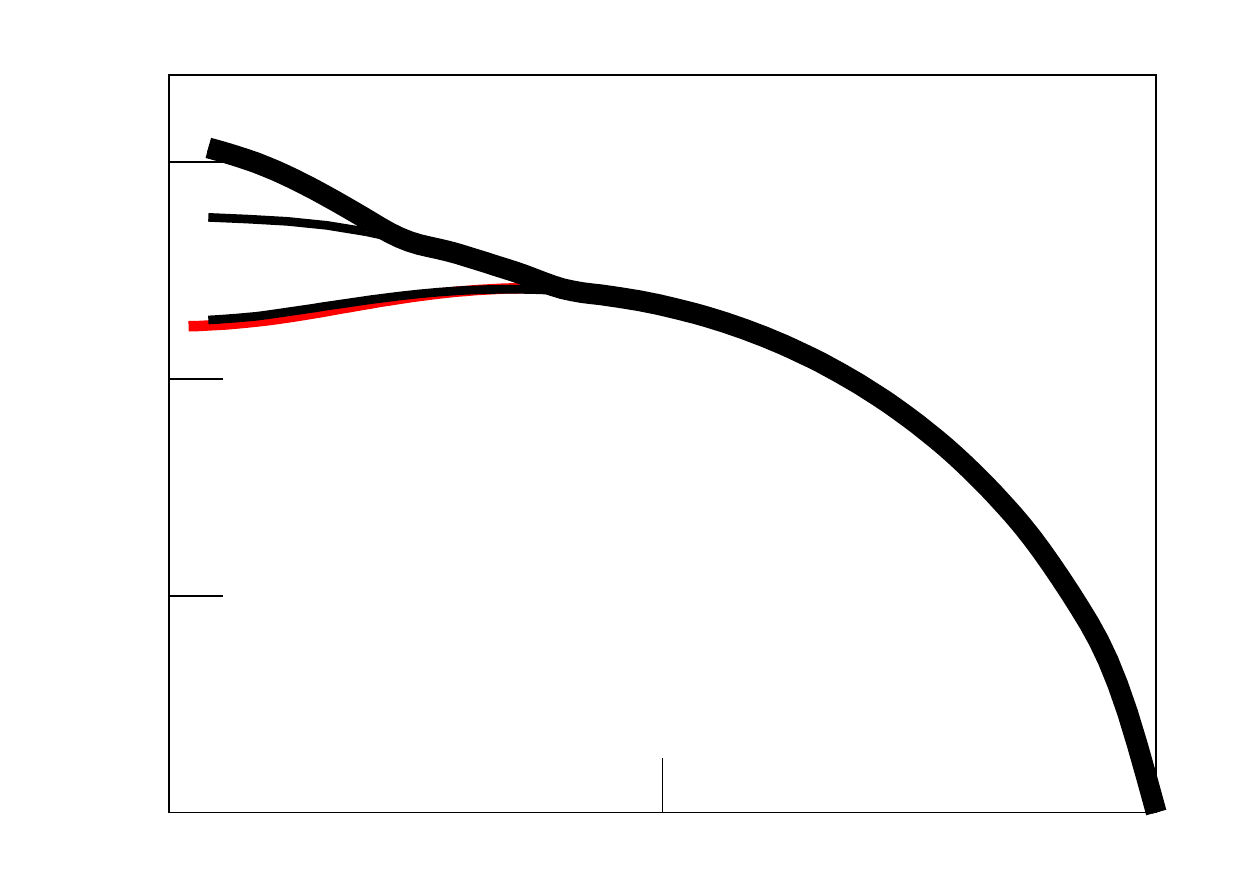} &
\includegraphics[width= \colwidth]{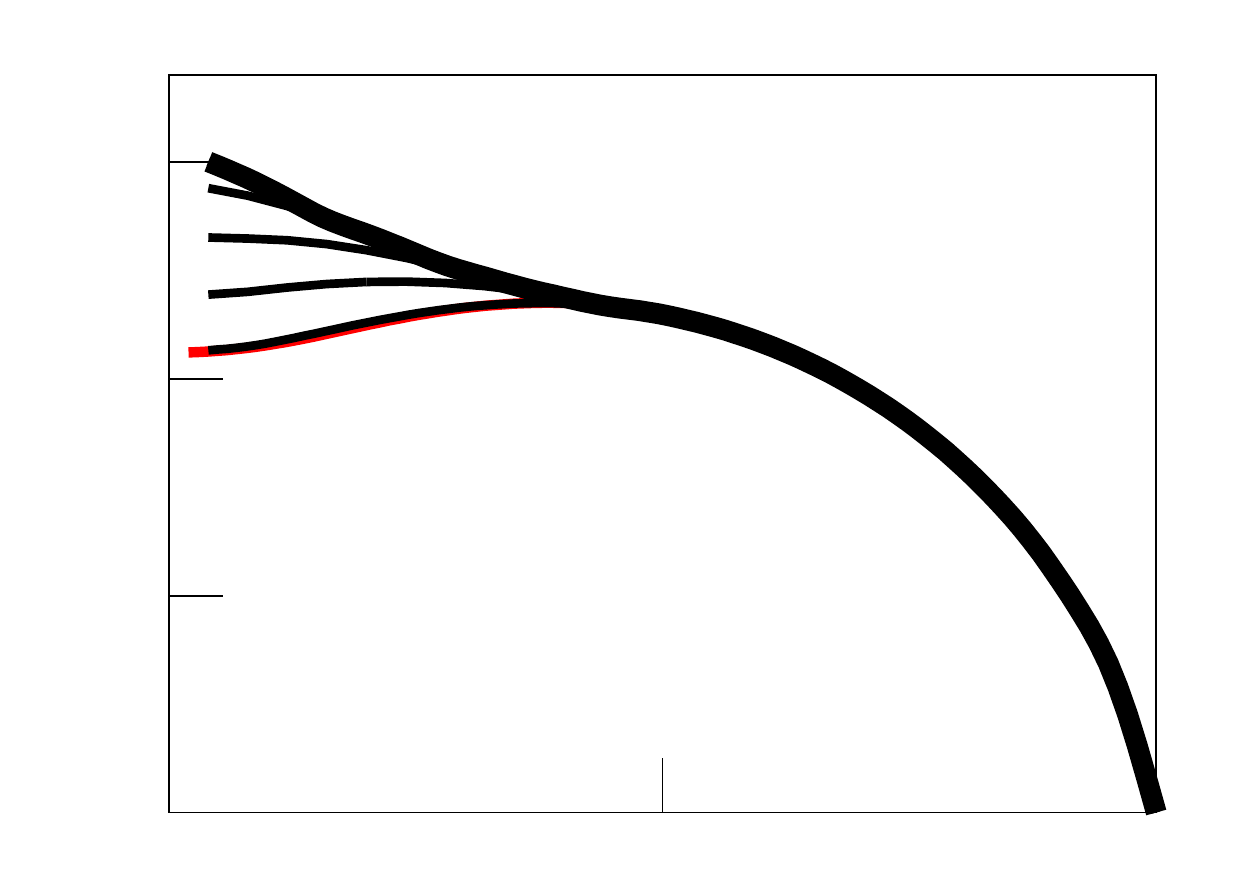} &
$\lambda_{so}=1/2\alpha$ \\
\includegraphics[width= \colwidth]{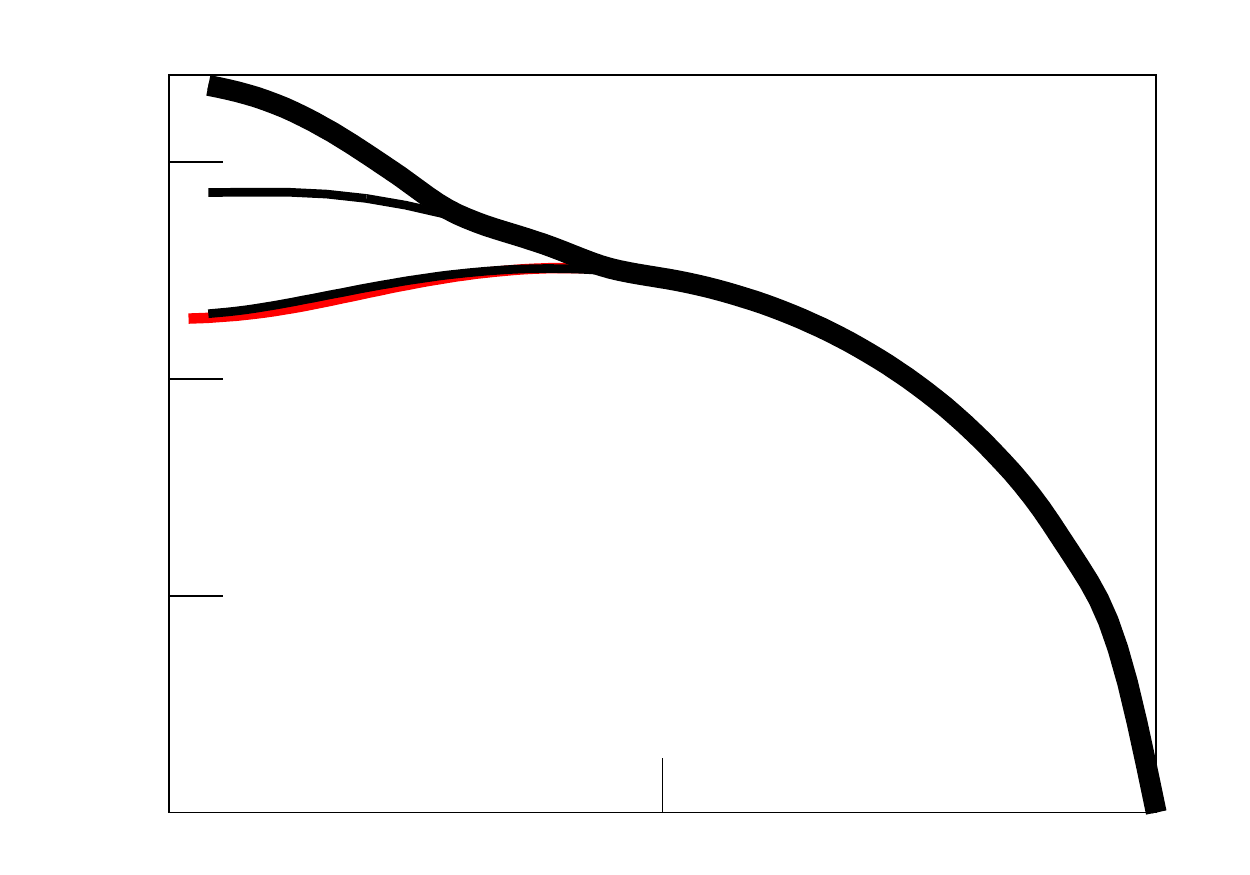} &
\includegraphics[width= \colwidth]{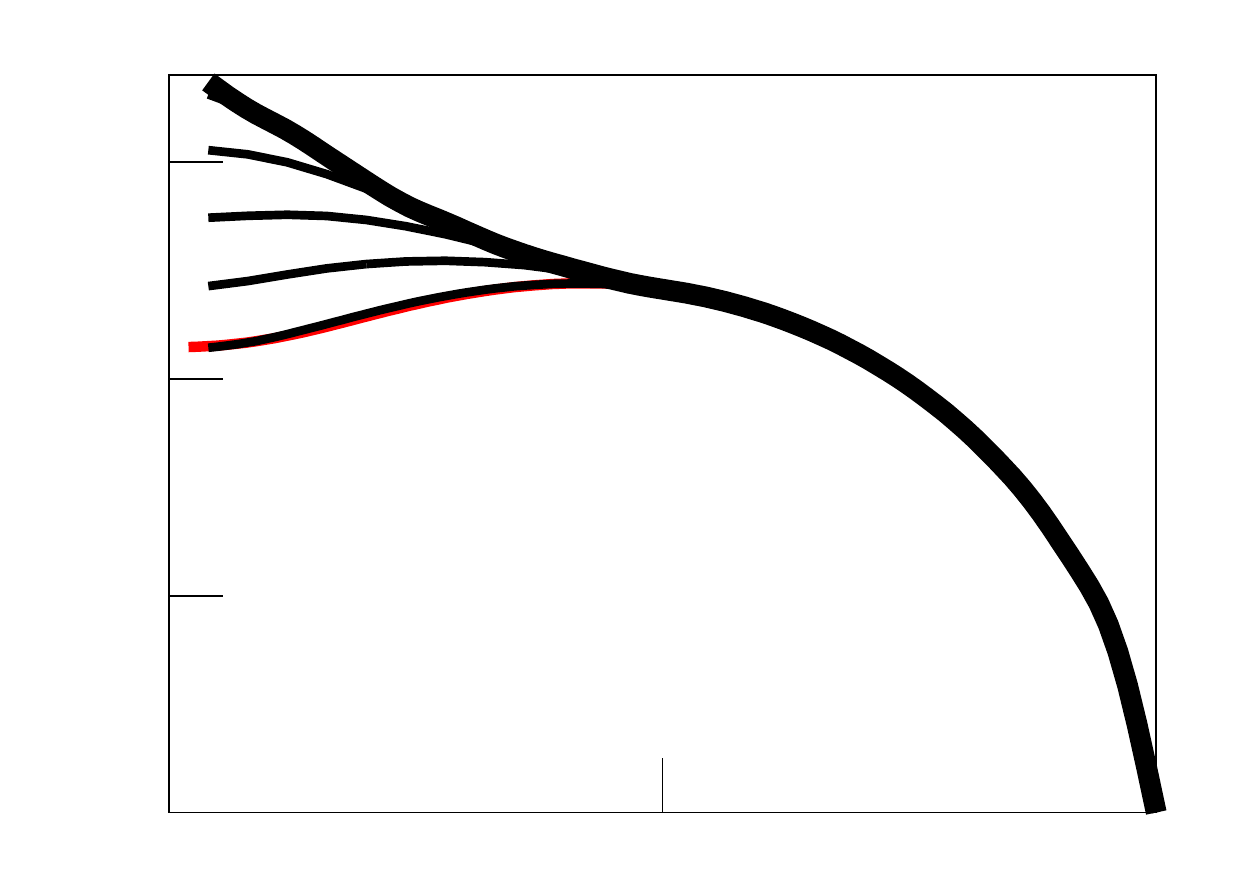} &
\includegraphics[width= \colwidth]{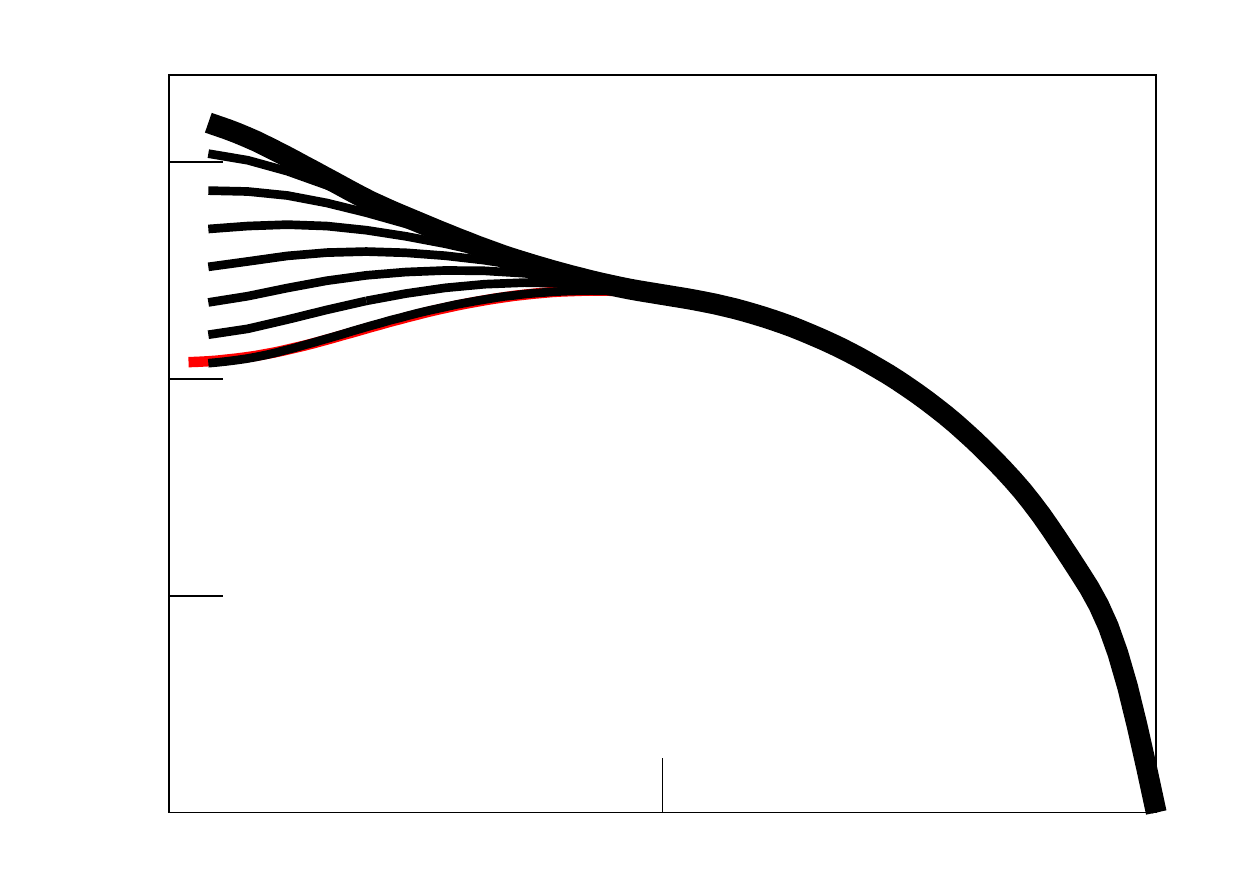} &
$\lambda_{so}=1/4\alpha$
\end{tabular*}\end{center}
\caption{\label{fig:fflophases}(Color online) The model exhibits a variety of phase diagram shapes. X-axis is temperature on the interval $[0,T_c]$. Y-axis is field, with the middle tick mark at $H_p$. The WHH solution appears in red, in agreement with our model's ground state.}
\end{figure}

\section{Experimental validation}
\begin{figure*}
\includegraphics[width=0.497\textwidth]{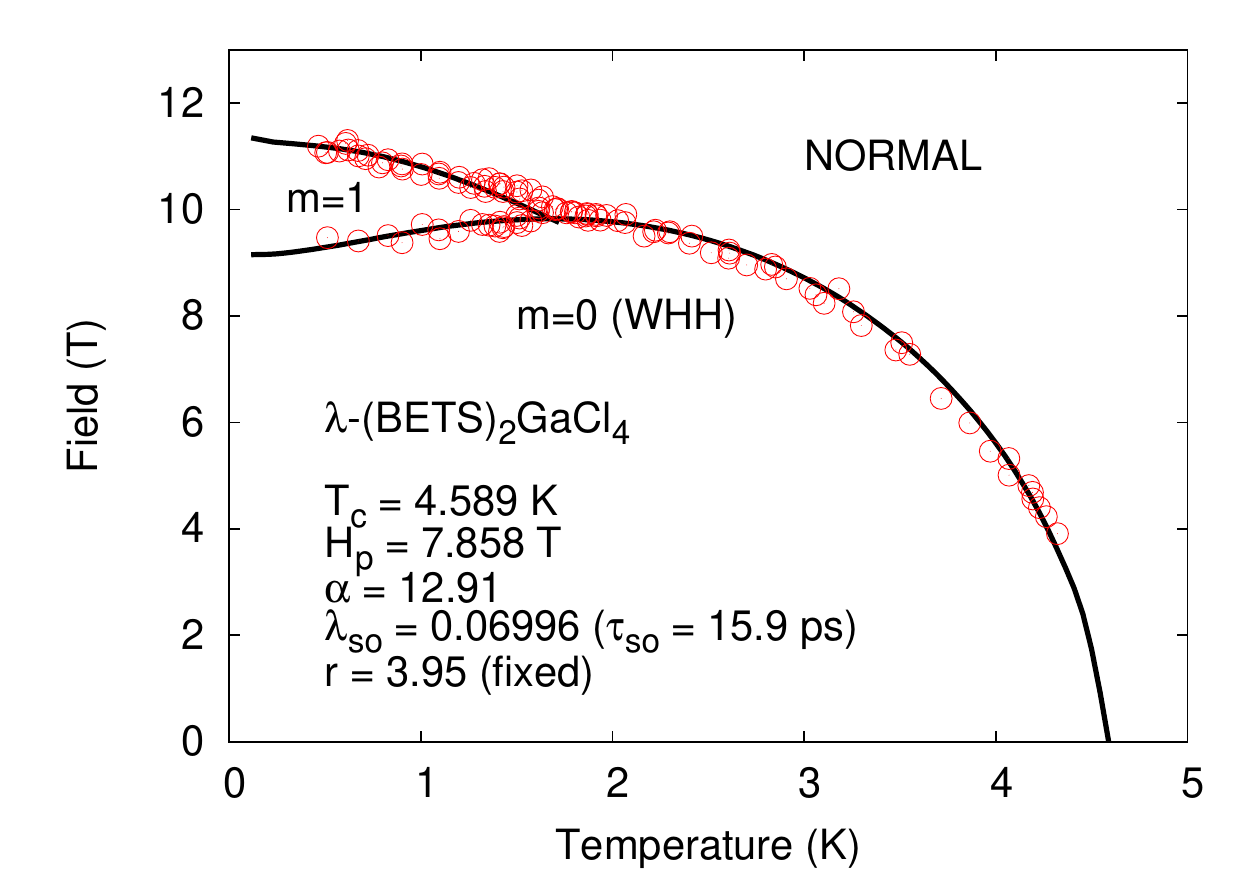}\includegraphics[width=0.497\textwidth]{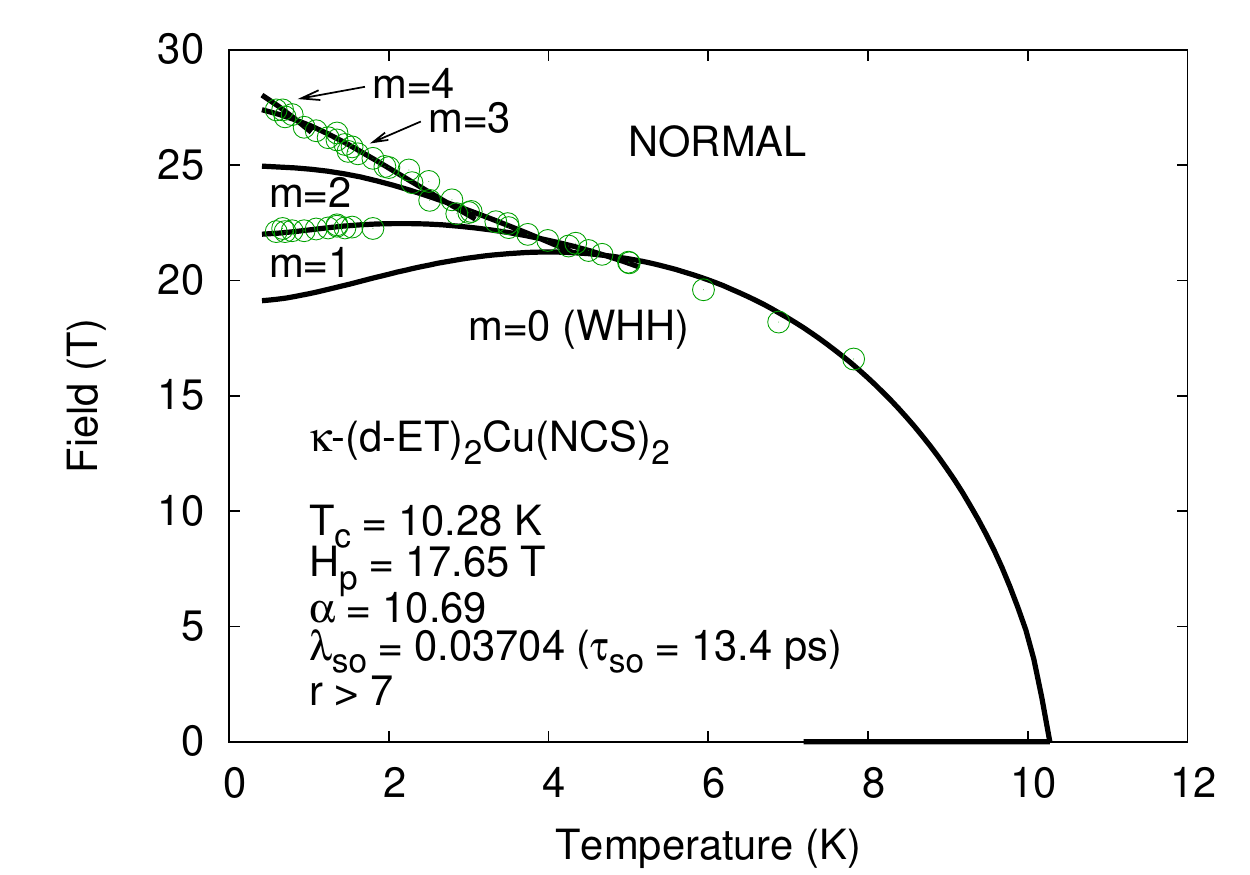}\\
\includegraphics[width=0.497\textwidth]{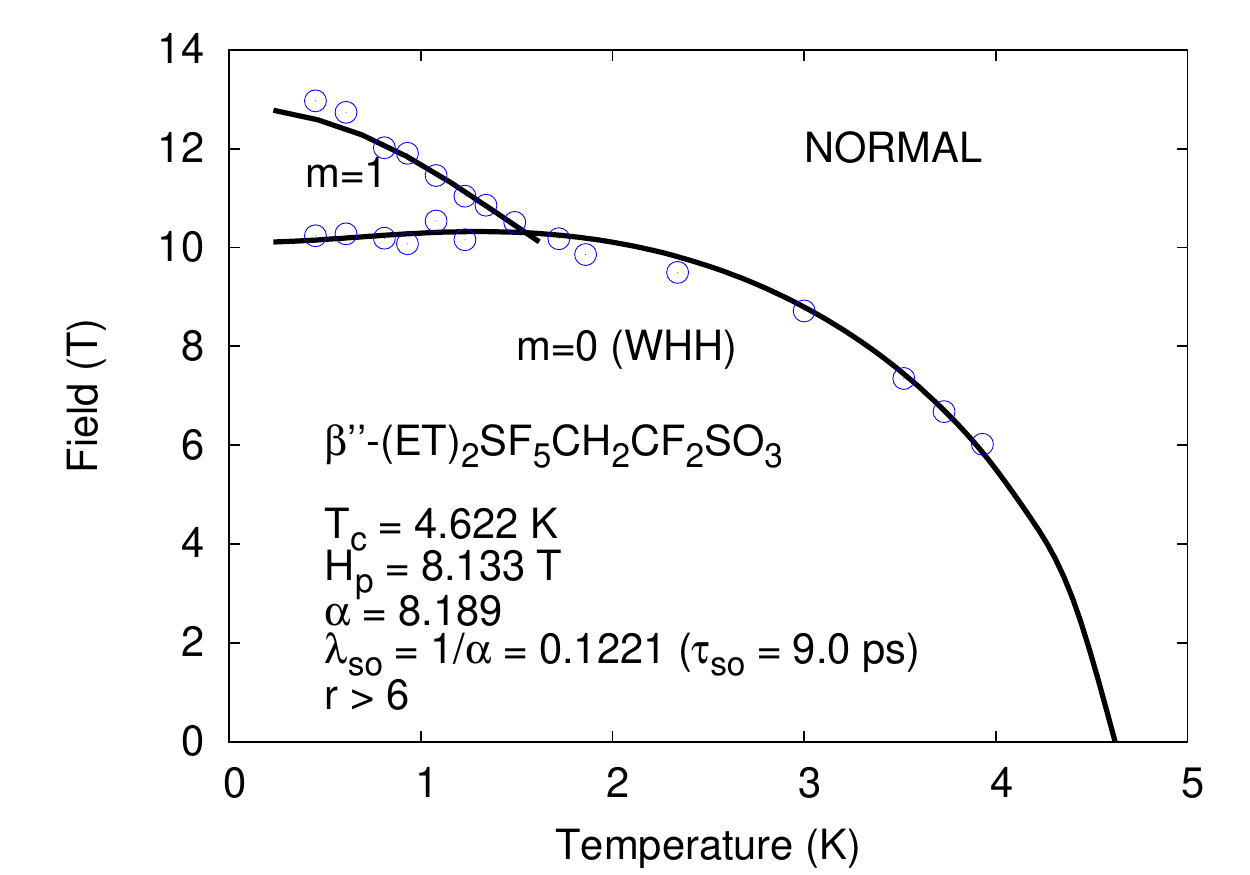}\includegraphics[width=0.497\textwidth]{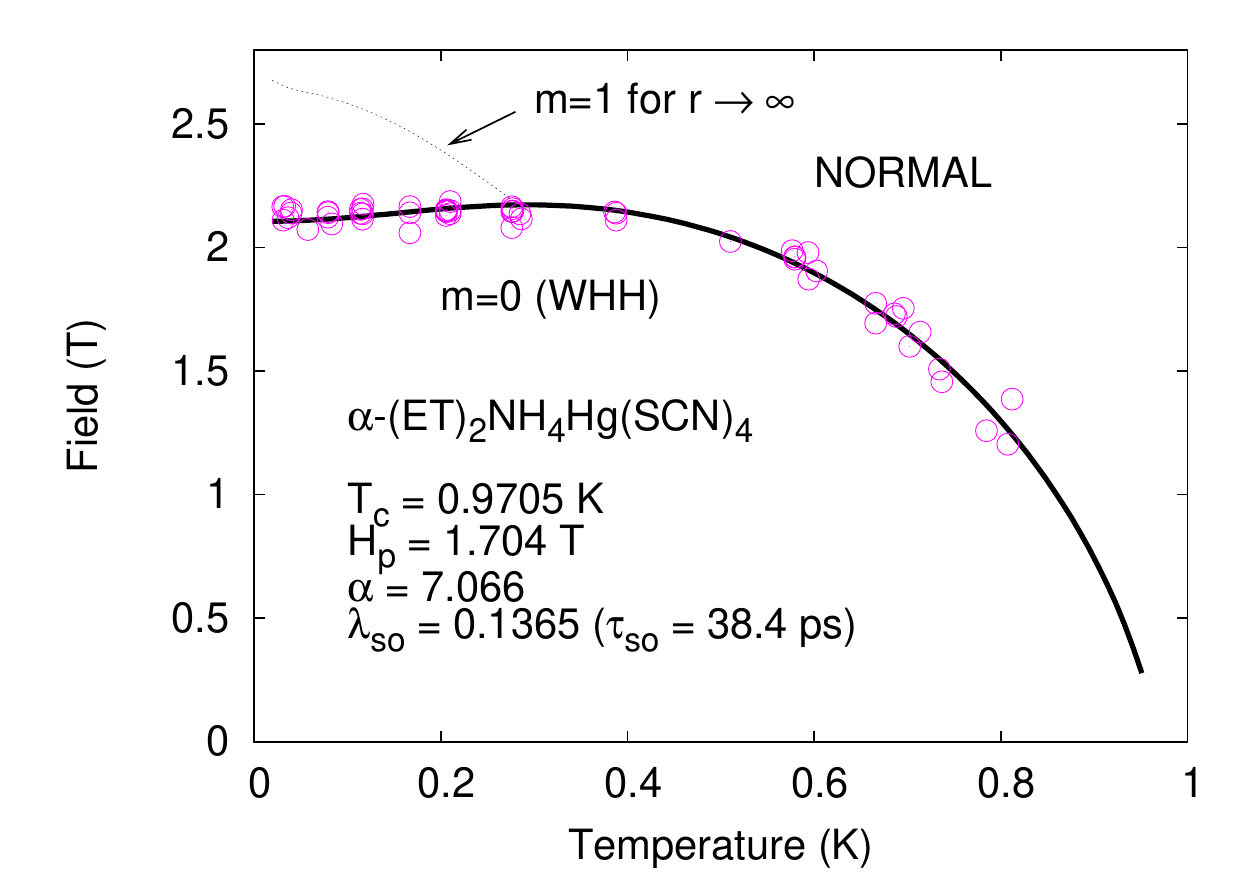}
\caption{\label{fig:experiment}(Color online) Phase diagram data and model fits for four organic superconductors. Top: \betsgacl\ and \cuncs. Bottom: \etpoly\ and \etnhfour.}
\end{figure*}
Experimental phase diagram data comes from a set of similar experiments on different samples using a Tunnel Diode Oscillator (TDO) as a penetration depth transducer. Penetration depth is preferred over resistivity because the latter tends to zero below $T_c$ and $H_{c2}$, while penetration depth reflects the Cooper pair density in different superconducting phases. Samples investigated were \betsgacl,\cite{coniglio_bets_2011} \etpoly,\cite{cho2009} \etnhfour,\cite{coffey_nhfour_2010} and \cuncs.\cite{agosta_cuncs} Of them, all but the \etnhfour\ exhibit FFLO behavior. The null result is particularly interesting, as it leads us to investigate why this very much paramagnetically limited superconductor shows no FFLO behavior.

We obtained fits to each data set by performing nonlinear least squares fitting using the Levenberg-Marquardt algorithm\cite{numericalrecipes} on the results of Eq.~\ref{eq:fflocleanlimit} with all data points and their respective phase diagram branches weighted equally.
%For each fit, we obtain the four WHH parameters along with the minimized goodness-of-fit metric $\sigma=\chi/N$, the deviation of data about the fit.
For each fit, we provide the parameters as well as the calculated spin-orbit scattering time $\tau_{so}$. As a practical modeling note, the magnetic field in WHH notation ($\bar{h}$) may be normalized to the Chandrasekhar-Clogston paramagnetic limit by $H_{c2}/H_p = \pi^2\alpha\bar{h} / 2.772$.\cite{coniglio_bets_2011}

%
%\begin{figure}
%\includegraphics[width=0.5\textwidth]{bets}
%\caption{\label{fig:bets}\betsgacl\ from \cite{coniglio_bets_2011}}
%\end{figure}
The FFLO phase in \betsgacl\ was observed as early as 2002 using thermal conductivity.\cite{tanatar_ishiguro_prb_2002} Here, using a 124 point phase diagram, we find that FFLO behavior is limited by the finite mean free path of the normal electrons, which prevents the higher field $m=2$ phase from forming in the given samples. In-plane anisotropy of the mean free path in a similar sample from the same batch\cite{mielke_singleton_2001} gives $r_x=3.5$ and $r_y=12$. Since the observed suppression of the  $m=2$ phase due to impurities occurs at $r<=3.95$, we conclude that for in-plane anisotropy in scattering, the more impure direction limits FFLO behavior. For a large computational penalty, the Laguerre polynomials in Eq.~\ref{eq:fflocleanlimit} could be expanded with an angular integration to investigate this behavior more carefully.

%\begin{figure}
%\includegraphics[width=0.5\textwidth]{etpoly}
%\caption{\label{fig:etpoly}\etpoly\ from \cite{cho2009}}
%\end{figure}
At the same $T_c$, we next consider \etpoly, an exceptionally clean superconductor with no metals in either the anion or cation.\cite{geiser_etsf5_1996} The FFLO phase rises sharply from the tricritical point, a feature our model reproduces under high spin-orbit scattering rates. The vortex products of spin-orbit scattering appear to be the sole source of orbital limiting. Preliminary unconstrained fitting gave $\lambda_{so}$ a very small negative value. Sine this corresponds to the unphysical possibility of orbitals enhancing rather than limiting $H_{c2}$, we must set the term under the square root, $1-\alpha\lambda_{so}$, to zero rather than accept the small negative value that would otherwise optimize the fit.

%\begin{figure}
%\includegraphics[width=0.5\textwidth]{nhfour}
%\caption{\label{fig:nhfour}\etnhfour\ from \cite{coffey_nhfour_2010}}
%\end{figure}
Coffey \textit{et.\ al.}\cite{coffey_nhfour_2010} find no FFLO phase in \etnhfour, despite a careful search, but they do see a very slightly reentrant phase line with a slightly lower critical field at low temperature than at 0.3 K. They report $r=\ell/\xi_0^\parallel\approx1.4$, which is well below the threshold (3.5) of FFLO behavior possible in our model. Correspondingly, we expect no FFLO phase. Upon fitting to the WHH model, we discovered normalized fit parameters very close to that of \etpoly. In particular, we again find $\lambda_{so}\approx1/\alpha$. %In the context of Eq.~\ref{eq:fflocleanlimit} or~\ref{eq:fflofull}, whose ground states ($m=0$) follow WHH theory, 
Both \etpoly\ and \etnhfour\ are known to be very anisotropic due to the thickness of their poorly conducting anion layers.\cite{wosnitza_etpoly_interlayer_trans_2002} Since vortices fit easily between the conducting layers, their contribution to orbital limiting is small. Almost all of the observed orbital limiting is therefore due to spin-orbit scattered vortices through the layers.

%\begin{figure}
%\includegraphics[width=0.5\textwidth]{cuncs}
%\caption{\label{fig:cuncs}\cuncs\ from \cite{agosta_cuncs}}
%\end{figure}
One of the most commonly studied organic superconductors is \cuncs. We examine a deuterated variant, although the difference appears to be only a slight increase in gap energy in the heavier compound. Our difficulties in modeling the observed behavior provide excellent opportunities to extend the theory and discover more variants of FFLO ordering. We were unable to match the transition at 22 T to the modeled $m=0$ phase line as worked so well for the other samples; however, it fits $m=1$ quite well. To our knowledge, there have been no reports of a phase boundary at 19 T where the fitted $m=0$ line appears. The broad superconducting transition makes a search for additional phase lines between 22 and 28 T challenging as well. With evidence pointing toward d-wave superconductivity\cite{agosta_cuncs} in \cuncs, it may be fruitful to incorporate pairing symmetry here, as has been done for other models.\cite{shimahara_rainer_1997} Alternatively, the radial symmetry proposed by Buzdin and Brison may not be correct for this compound.

\section{Discussion}
%bets 1.71 
%cuncs 1.72
%poly 1.76
%nh4 1.76
It is remarkable that all four samples for which we have phase diagram data realize a fit by this model, despite differences in their shape. Perhaps more remarkable, a comparison of their properties reveals a number of expected and unexpected results. The effect of $\alpha$ and $r$ on the FFLO state confirms the common hypothesis\cite{agosta_martin06_jopcs} that the search for paramagnetically limited and FFLO superconductors must focus on the $\alpha\gg 1$ and $r>3.5$ space. Large $\alpha$ is obtained by choosing quasi-2d compounds that exhibit the vortex lock-in effect or heavy fermion superconductors, and $\alpha$ can be estimated from the slope of $H_{c2}$ near $T_c$ without the need for particularly high field magnets. Large $r$ comes from choosing compounds that can either be grown crystallographically pure or those whose coherence length can be made very small, again favoring the organics and heavy fermion class.

Long range ordering is necessary for FFLO behavior. In this case, the electronic mean free path must be larger than the vortices of Fig.~\ref{fig:vortices}. We parameterize the cleanliness of the material by the normalized mean free path $r$, which is most easily measured in high fields by rotating the sample $90^\circ$ and measuring the Dingle temperature via Shubnikov de Haas oscillations. Oscillations in \etpoly\ and \cuncs\ are both readily observable and yield $r>10$, as expected from our model fit. Oscillations in \etnhfour\ are observable, but superconductivity is so weak that $r\approx 1$, in line with the absence of FFLO behavior. \betsgacl\ is an interesting case, as $r=3.5$ from measurements,\cite{mielke_singleton_2001} and this model requires $r=3.95$ to fit the phase diagram and avoid the appearance of an $m=2$ state, which was not observed. Such error is reasonable for a comparison of crystal purity, and supports our model's ability to account for electronic mean free path numerically in the phase diagram.

Finally, we report a surprising constant relationship. Taking $H_p/T_c$ from Fig.~\ref{fig:experiment} gives values, 1.71, 1.72, 1.76, and 1.76. Although the BCS theory of superconductivity does not necessarily apply to the organics, BCS gives $H_p/T_c=1.83$ as a constant. Attempts have been made to calculate values for the organics\cite{agosta_cuncs} but have lacked definite values for $H_p$, which we observe is not the same as $H_{c2}(T=0)$. If $H_p/T_c\approx1.74$ is really a constant, that is quite a useful value in the search for a microscopic pairing mechanism in these superconductors.

We have developed an FFLO phase diagram model based on theory by Buzdin and Brison and show that this model may be applied with remarkable success to the observed FFLO phase in several organic superconductors. From the model, we learn that the FFLO ordering in the organics is radial instead of plane wave. From the fits, we confirm hypotheses and learn relationships such as anisotropy vs.\ shape of FFLO phase diagram, crystal purity vs.\ phase diagram, and superconducting energy gap vs.\ Chandrasekhar-Clogston (Pauli) paramagnetic limit. We hope our work will spur the scientific community toward developing a greater understanding of superconductivity, paramagnetic limiting, and critical field enhancement through long-range ordering.

\section{Acknowledgements}
We appreciate the support of the U.S. Department of Energy, NSA SSAA DE-NA0001979 and ER46214. 

\bibliography{bibs/bets,bibs/organics,bibs/tdo,bibs/computation}{}

\end{document}